# Deep learning for Stock Market Prediction

**Mojtaba Nabipour [1], Pooyan Nayyeri [2], Hamed Jabani [3], Amir Mosavi [4,5,6,*]**

1. Faculty of Mechanical Engineering, Tarbiat Modares University, Tehran, Iran.
2. School of Mechanical Engineering, College of Engineering, University of Tehran, Tehran, Iran.
3. Department of Economics, Payame Noor University, West Tehran Branch, Tehran, Iran.
4. Institute of Structural Mechanics (ISM), Bauhaus-Universität Weimar, 99423 Weimar, Germany.
5. School of the Built Environment, Oxford Brookes University, Oxford OX30BP, UK.
6. Faculty of Civil Engineering, Technische Universität Dresden, 01069 Dresden, Germany.
* Correspondence: a.mosavi@brookes.ac.uk

**Abstract:** Prediction of stock groups values has always been attractive and challenging for shareholders. This paper concentrates on the future prediction of stock market groups. Four groups named diversified financials, petroleum, non-metallic minerals and basic metals from Tehran stock exchange are chosen for experimental evaluations. Data are collected for the groups based on ten years of historical records. The values predictions are created for 1, 2, 5, 10, 15, 20 and 30 days in advance. The machine learning algorithms utilized for prediction of future values of stock market groups. We employed Decision Tree, Bagging, Random Forest, Adaptive Boosting (Adaboost), Gradient Boosting and eXtreme Gradient Boosting (XGBoost), and Artificial neural network (ANN), Recurrent Neural Network (RNN) and Long short-term memory (LSTM). Ten technical indicators are selected as the inputs into each of the prediction models. Finally, the result of predictions is presented for each technique based on three metrics. Among all algorithms used in this paper, LSTM shows more accurate results with the highest model fitting ability. Also, for tree-based models, there is often an intense competition between Adaboost, Gradient Boosting and XGBoost.

## 1. Introduction

The prediction process of stock values is always a challenging problem [1] because of its unpredictable nature. The dated market hypothesis believe that it is impossible to predict stock values and that stocks behave randomly, but recent technical analyses show that the most stocks values are reflected in previous records, therefore the movement trends are vital to predict values effectively [2]. Moreover, stock market's groups and movements are affected by several economic factors such as political events, general economic conditions, commodity price index, investors' expectations, movements of other stock markets, psychology of investors, etc [3]. The value of stock groups is computed with high market capitalization. There are different technical parameters to obtain statistical data from value of stocks prices [4]. Generally, stock indices are gained from prices of stocks with high market investment and they often give an estimation of economy status in each country. For example, findings prove that economic growth in countries is positively impacted by the stock market capitalization [5].

The nature of stock values movement is ambiguous and makes investments totally risky for investors. Also, it is usually a big problem to detect the market status for governments. It is true that the stock values are generally dynamic, non-parametric and non-linear; therefore they often cause weak performance of the statistical models and disability to predict the accurate values and movements [6, 7]

Machine learning is the most powerful tool which includes different algorithms to effectively develop their performance on a certain case study. It is common belief that ML have a significant ability of identifying valid information and detecting patterns from the dataset [8].

In contrast with the traditional methods in the ML area, the ensemble models are a machine learning based way in which some common algorithms are used to work out a particular problem, and have been confirmed to outperform each of methods when predicting time series [9-11]. For prediction problems in machine learning area, boosting and bagging are effective and popular algorithms among ensemble ways. There is recent progress of tree based models with introducing gradient boosting and XGBoost algorithms, which have been significantly employed by top data scientists in competitions. Indeed, a modern trend in ML, which is named deep learning (DL), can deem a deep nonlinear topology in its specific structure, has its excellent ability from the financial time series to extract relevant information [12]. Contrary to simple artificial neural network, recurrent neural networks (RNN) have achieved a considerable success in the financial area on account of their great performance [13, 14]. It is clear that the prediction process of the stock market is not only related to the current information but the earlier data has a vital role, so the training will be insufficient if only the data is used at the latest time. RNN is able to employ the network to sustain memory of recent events and build connections between each unit of a network, so, it is completely proper for the economic predictions [15, 16]. Long short-term memory (LSTM) is an improved subset of RNN method which used in deep learning area. LSTM has three different gates to remove the problems in RNN cells and also is able to process single data points or whole sequences of data.

In academic fields, many studies have been conducted on market prediction ways. Also, there are various approaches to time series modeling. Exponential smoothing ,moving average and ARIMA are common linear models for predicting future prices [17, 18]. Several research activities have done for extensive predictions with Artificial Neural Networks (ANN), Genetic Algorithms (GA), fuzzy logic etc [19-21]. Zhang et al. [22] combined Improved Bacterial Chemotaxis Optimization (IBCO) with artificial neural network. They indicated that their proposed method is able to predict stock index for a short time (1 day ahead) and a long time (15 days ahead), and their outcomes showed the excellent results of the method. Asadi et al. [1] used preprocessing ways as a combination of data, by feed forward neural networks and employing genetic algorithms and Levenberg–Marquardt (LM) method for learning. Preprocessing ways such as data transformation and selection of input variables were employed for developing the model performance. The final results demonstrated that the proposed method was capable of dealing with the stock market fluctuations with suitable prediction accuracy. Shen, Guo, Wu et al [23] introduced the Artificial Fish Swarm Algorithm (AFSA) for training radial basis function neural network (RBFNN). Their experimental works was based on data from Shanghai Stock Exchange to show that the optimized RBF by AFSA was a practical method with significant accuracy. Jigar et al. [24] predicted the Indian stock market index by a combination of machine learning methods; they considered two different stages, a single stage scenario in comparison with hybrid combination of models with better results. S Olaniyi et al [25] supposed a linear regression method of analyzing stock market behaviors. The approach successfully predicted stock prices based on two parameters.

This study concentrates on the process of future values prediction for stock market groups, which are totally crucial for investors. The predictions are evaluated for 1,2,5,10,15,20 and 30 days in advance. It has been noted from the research background, the most of them focused on classification problems instead of regression ones [26-28]. By considering literature review, this research work examines the prediction performance of a set of cutting-edge machine learning methods, which involves tree-based models and neural networks. Also, employing the whole of tree-based methods, RNN and LSTM techniques for regression problems in the stock market area is a novel research activity which presented in this study.

This paper involves three different sections. At the first, through methodology section, the evolution of tree-based models with the introduction of each one are presented. In addition, basic structure of neural networks and recurrent ones are described briefly. In the research data section, ten technical indicators are shown in detail with selected methods parameters. At the final step, after introducing three regression metrics, machine learning results are reported for each group, and the models behavior are compared.

## 2. Materials and Methods

2.1. Tree-based models

Since the set of splitting rules employed to differently divide the predictor space can be summarized in a tree, these types of models are known as decision-tree methods. Fig 1 shows the evolution of tree-based algorithms over several years.

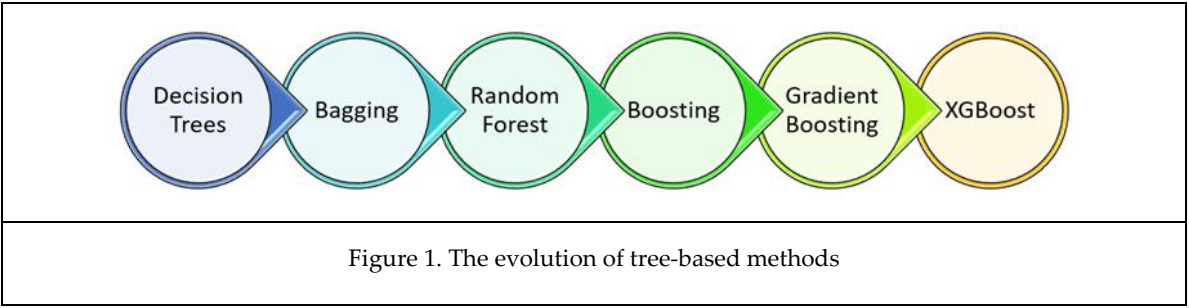

Figure 1. The evolution of tree-based methods

2.1.1. Decision Tree

Decision Trees are a popular supervised learning technique used for classification and regression jobs. The purpose is to make a model that predicts a target value by learning easy decision rules formed from the data features. There are some advantages of using this method like being easy to understand and interpret or Able to work out problems with multi-outputs; on the contrary, creating over-complex trees which results in overfitting is a fairly common disadvantage. A schematic illustration of Decision tree is shown in Fig 2.

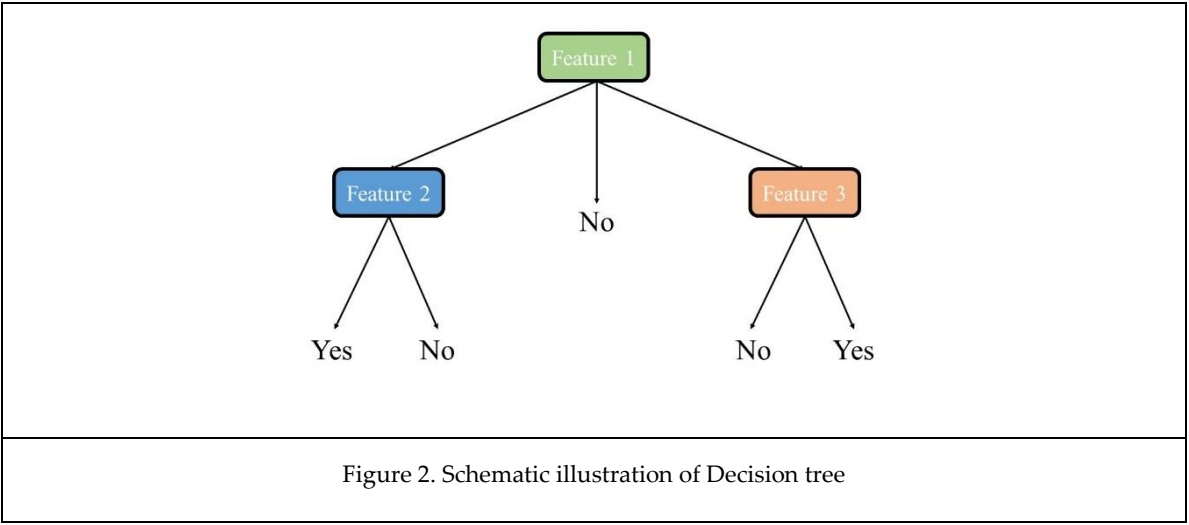

Figure 2. Schematic illustration of Decision tree

2.1.2. Bagging

A Bagging model (as a regressor model) is an ensemble estimator that fits each basic regressor on random subsets of the dataset and next accumulate their single predictions, either by voting or by averaging, to make the final prediction. This method is a meta-estimator and can commonly be employed as an approach to decrease the variance of an estimator like a decision tree by using randomization into its construction procedure and then creating an ensemble out of it. In this method samples are drawn with replacement and predictions, and obtained through a majority voting mechanism.

2.1.3. Random Forest

The random forest model is created by great number of decision trees. This method simply averages the prediction result of trees, which is called a forest. Also, this model has three random concepts, randomly choosing training data when making trees, selecting some subsets of features when splitting nodes and considering only a subset of all features for splitting each node in each simple decision tree. During training data in a random forest, each tree learns from a random sample of the data points. A schematic illustration of Random forest is indicated in Fig 3.

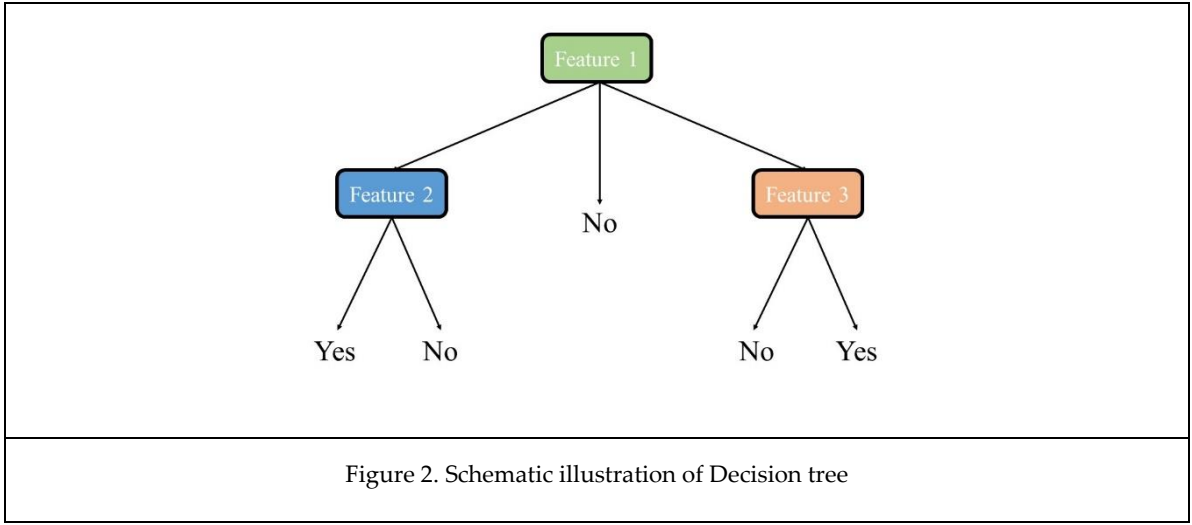

Figure 2. Schematic illustration of Decision tree

2.1.4. Boosting

Boosting method refers to a group of algorithms which converts weak learners to a powerful learner. The method is ensemble for developing the model predictions of any learning algorithm. The concept of boosting is to sequentially train weak learners in order to correct its past performance. AdaBoost is a meta-estimator that starts by fitting a model on the main dataset and then fits additional copies of the model on the similar dataset. During the process, samples' weights are adapted based on the current prediction error, so subsequent models concentrates more on difficult items.

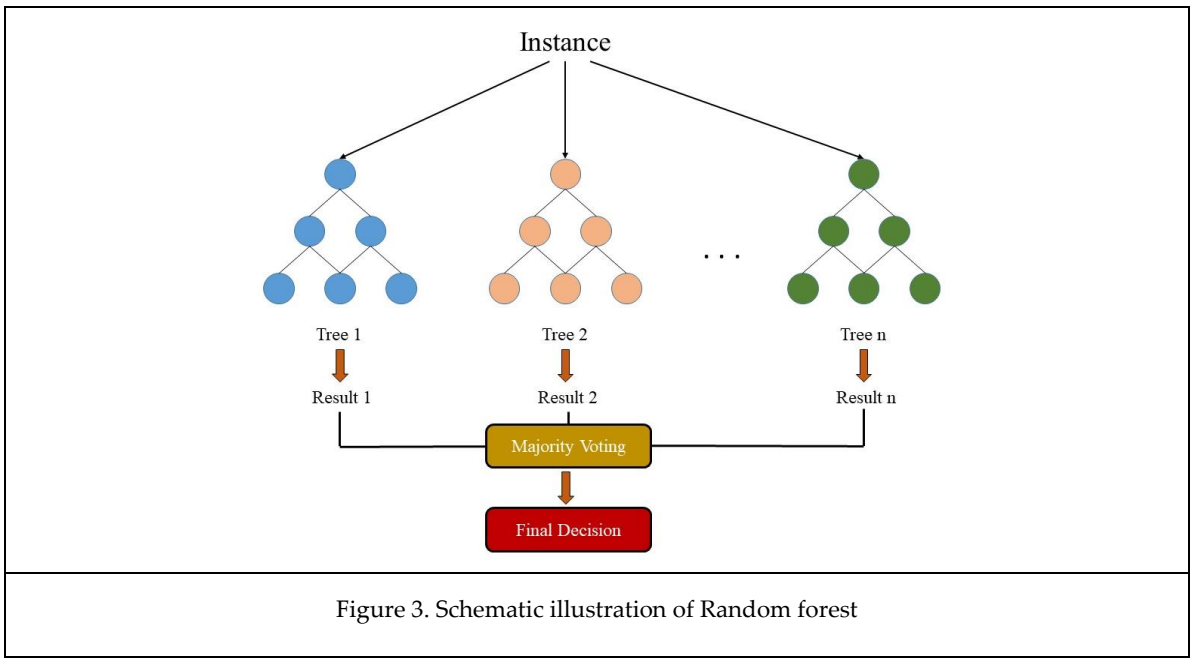

Figure 3. Schematic illustration of Random forest

2.1.5. Gradient Boosting

Gradient Boosting method is like AdaBoost when it sequentially adds predictors to an ensemble model, each of them corrects its past performance. In contrast with AdaBoost, Gradient Boosting fits a new predictor to the residual errors (made by the prior predictor) with using gradient descent to find the failing in the predictions of previous learner. Overall, the final model is capable of employing for the base model to decreases errors over the time.

2.1.6. XGBoost

XGBoost is an ensemble tree method (like Gradient Boosting ) and the method apply the principle of boosting for weak learners. However, XGBoost was introduced for better speed and performance. In-built cross-validation ability, efficient handling of missing data, regularization for avoiding overfitting, catch awareness, tree pruning and parallelized tree building are common advantages of XGBoost algorithm.

2.2. Artificial neural networks

2.2.1. ANN

ANN are single or multi-layer neural nets which fully connected together. Fig 4 shows a sample of ANN with an input and ouput layer and also two hidden layers. In a layer, each node is connected to every other node in the next layer. By increase in the number of hidden layers, it is possible to make the network deeper. A schematic illustration of ANN is demonstrated in Fig 4.

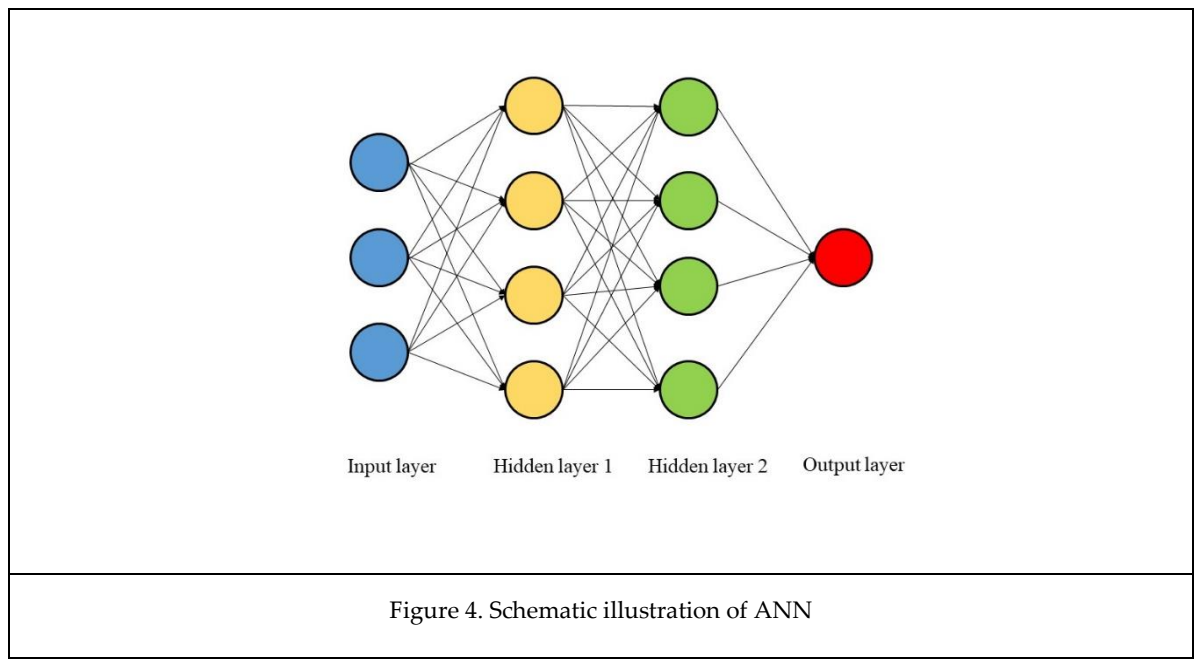

Figure 4. Schematic illustration of ANN

Fig 5 is shown for each of the hidden or output nodes, while a node takes the weighted sum of the inputs, added to a bias value, and passes it through an activation function (usually a non-linear function). The result is the output of the node that becomes another node input for the next layer. The procedure moves from the input to the output, and the final output is determined by doing this process for all nodes. Learning process of weights and biases associated with all nodes for training the neural network.

The Equation 1 shows the relationship between nodes and weights, and biases [29]. The weighted sum of inputs for a layer passed through a non-linear activation function to another node in the next layer. It can be interpreted as a vector, where $X_1$, $X_2$ ... and $X_n$ are inputs, $w_1$, $w_2$, ... and $w_n$ are weights respectively, n is the inputs number for the final node, f is activation function and z is the output.

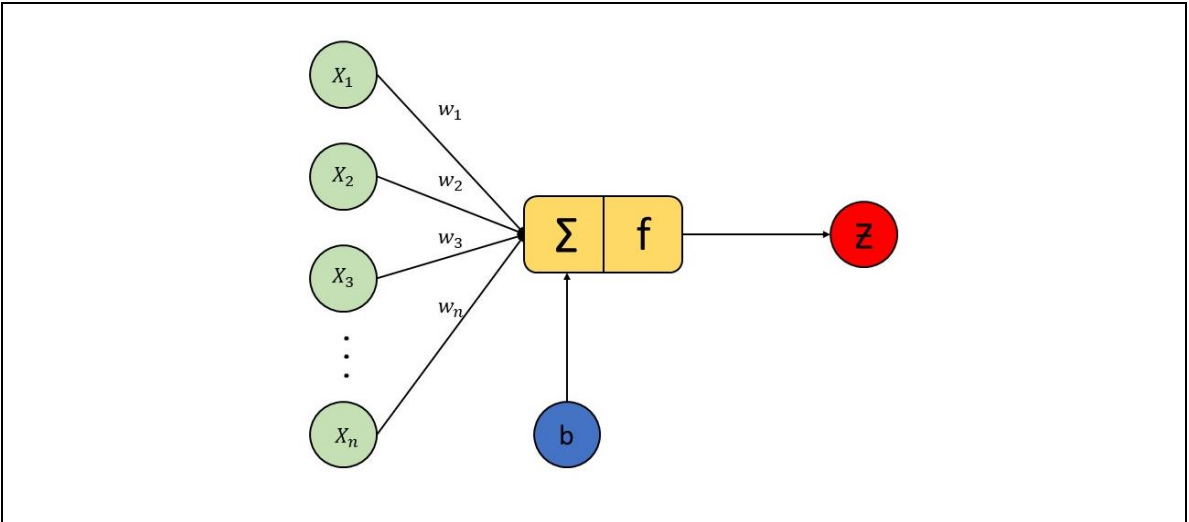

Figure 5. An illustration of relationship between inputs and output for ANN

$$Z = f(x.w + b) = f\left(\sum_{i=1}^{n} x_i w_i + b\right) \tag{1}$$

By calculating weights/biases, the training process is completed by some rules: initialize the weights/biases for all the nodes randomly, performing a forward pass by the current weights/biases and calculating each node output, comparing the final output with the actual target, and modifying the weights/biases consequently by gradient descent with backwards pass, generally known as backpropagation algorithm.

2.2.2. RNN

RNN is a very prominent version of neural networks extensively used in various processes. In a common neural network, an input is processed through a number of layers and an output is made. It is assumed that two consecutive inputs are independent of each other. However, the situation is not correct in all processes. For example, for prediction of the stock market at a certain time, it is crucial to consider the previous observations.

RNN is named recurrent due to it does the same task for each item of a sequence when the output is related to the previous computed values. As another important point, RNN has a specific memory, which stores previous computed information for a long time. In theory, RNN can use information randomly for long sequences, but in real practices, there is a limitation to look back just a few steps. Fig 6 shows the architecture of RNN.

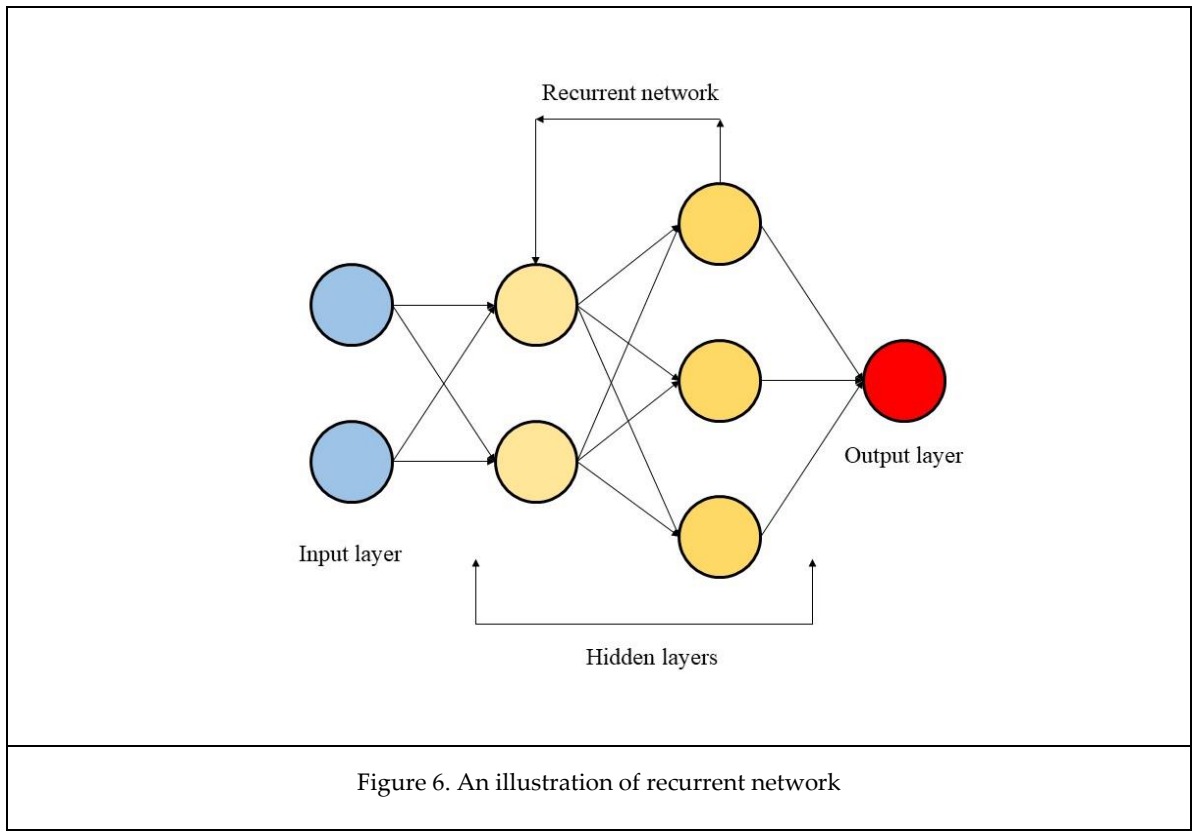

Figure 6. An illustration of recurrent network

### 2.2.3. LSTM

LSTM is a specific kind of RNN with wide range of applications like time series analysis, document classification, speech and voice recognition. In contrast with feedforward ANNs, the predictions made by RNNs are dependent on previous estimations. In real, RNNs are not employed extensively because they have a few deficiencies which cause impractical evaluations.

Without investigation of too much detail, LSTM solves the problems by employing assigned gates for forgetting old information and learning new ones. LTSM layer is made of four neural network layers that interact in a specific method. A usual LSTM unit involves three different parts, a cell, an output gate and a forget gate. The main task of cell is recognizing values over random time intervals and the task of controlling the information flow into the cell and out of it belongs to the gates.

### 3. Research data

This paper employs data from November 2009 to November 2019 (ten years) of four stock market groups, Diversified Financials, Petroleum, Non-metallic minerals and Basic metals, which are completely generous. From opening, close, low high and prices of the groups, ten technical indicators are calculated. The whole of data for the study is acquired from www.tsetmc.com website. As an important point, to prevent the effect of the larger value of an indicator on the smaller one, the values of ten technical indicators for all groups are normalized independently. Table 1 indicate all the technical indicators, which are employed as input values.

Table 1. Selected Technical Indicators (n is 10 here)

Simple n-day moving average = $\frac{C_t + C_{t-1} + \cdots + C_{t-n+1}}{n}$

Weighted 14-day moving average = $\frac{n*C_t + (n-1)*C_{t-1} + \cdots + C_{t-n+1}}{n + (n-1) + \cdots + 1}$

Momentum = $C_t - C_{t-n+1}$

Stochastic K% = $\frac{C_t - LL_{t\_t-n+1}}{HH_{t\_t-n+1} - LL_{t\_t-n+1}} * 100$

Stochastic D% = $\frac{K_t + K_{t-1} + \cdots + K_{t-n+1}}{n} * 100$

Relative strength index (RSI) = $100 - \frac{100}{1 + \sum_{i=1}^{n-1} UP_{t-i} / \sum_{i=1}^{n-1} DW_{t-i}}$

Signal(n)$_t$ = MACD$_t$ * $\frac{2}{n+1}$ + Signal(n)$_{t-1}$ * $(1 - \frac{2}{n+1})$

Larry William's R% = $\frac{HH_{t\_t-n+1} - C_t}{HH_{t\_t-n+1} - LL_{t\_t-n+1}} * 100$

Accumulation/Distribution (A/D) oscillator: $\frac{Ht - Ct}{Ht - Lt}$

CCI (Commodity channel index) = $\frac{M_t - SM_t}{0.015 D_t}$

While:

$C_t$ is the closing price at time t

$L_t$ and $H_t$ is the low price and high price at time t respectively

$LL_{t\_t-n+1}$ and $HH_{t\_t-n+1}$ is the lowest low and highest high prices in the last n days respectively

$UP_t$ and $DW_t$ means upward price change and downward price change at time t respectively

EMA(K)$_t$ = EMA(K)$_{t-1}$ * $(1 - \frac{2}{k+1})$ + $C_t * \frac{2}{k+1}$

Moving average convergence divergence (MACD$_t$) = EMA(12)$_t$ - EMA(26)$_t$

$M_t = \frac{Ht + Lt + Ct}{3}$

$SM_t = \frac{\sum_{i=0}^{n-1} M_{t-i}}{n}$

$D_t = \frac{\sum_{i=0}^{n-1} |M_{t-i} - SM_t|}{n}$

Dataset used for all models -except RNN and LSTM models- are identical. There are 10 features (10 technical indicators) and one target (stock index of the group) for each sample of the dataset. As mentioned, all 10 features are normalized independently before using to fit models to improve the performance of algorithms.

Since the goal is to develop models to predict stock group values, datasets are rearranged to incorporate the 10 features of each day to the target value of n-days ahead. In this study, models are evaluated by training them for predicting the target value for 1, 2, 5, 10, 15, 20, and 30 days ahead.

There are several parameters related each model. For tree-based models, number of trees (ntrees) is the design parameter while other common parameters are set identical between all models. Parameters and their values for each model are listed in Table 2.

Table 2. Tree-based Models parameters

| Model | Parameters | Value(s) |
|---|---|---|
| Decision Tree | Number of Trees (ntrees) | 1 |
| Bagging | Number of Trees (ntrees) | 50, 100, 150, 200, 250, 300, 350, 400, 450, 500 |
| | Max Depth | 10 |
| Random Forest | Number of Trees (ntrees) | 50, 100, 150, 200, 250, 300, 350, 400, 450, 500 |
| | Max Depth | 10 |
| Adaboost | Number of Trees (ntrees) | 50, 100, 150, 200, 250, 300, 350, 400, 450, 500 |
| | Max Depth | 10 |
| | Learning Rate | 0.1 |
| Gradient Boosting | Number of Trees (ntrees) | 50, 100, 150, 200, 250, 300, 350, 400, 450, 500 |
| | Max Depth | 10 |
| | Learning Rate | 0.1 |
| XGBoost | Number of Trees (ntrees) | 50, 100, 150, 200, 250, 300, 350, 400, 450, 500 |
| | Max Depth | 10 |
| | Learning Rate | 0.1 |

For RNN and LSTM networks, because of their time series behavior, datasets are arranged to include the features of more than just one day. While for ANN model all parameters but epochs are constant, for RNN and LSTM models the variable parameters are number of days included in training dataset and respective epochs. By increasing the number of days in training set, the number of epochs is increased to train the models with an adequate number of epochs. Table 3 presents all valid values for parameters of each model. For example, if 5 days are included in the training set for ANN or LSTM models, the number of epochs is set to 300 in order to thoroughly train the models.

Table 3. Neural Network Based Models parameters

| Model | Parameters | Value(s) |
|---|---|---|
| ANN | Number of Neurons | 500 |
| | Activation Function | *Relu* |
| | Optimizer | Adam ($\beta_1 = 0.9, \beta_2 = 0.999$) |
| | Learning Rate | 0.01 |
| | Epochs | 100, 200, 500, 1000 |
| RNN | Number of Neurons | 500 |
| | Activation Function | *tanh* |
| | Optimizer | Adam ($\beta_1 = 0.9, \beta_2 = 0.999$) |
| | Learning Rate | 0.0001 |
| | Training Days (ndays) | 1, 2, 5, 10, 20, 30 |
| | Epochs (w.r.t. ndays) | 100, 200, 300, 500, 800, 1000 |
| LSTM | Number of Neurons | 200 |
| | Activation Function | *tanh* |
| | Optimizer | Adam ($\beta_1 = 0.9, \beta_2 = 0.999$) |
| | Learning Rate | 0.0005 |
| | Training Days (ndays) | 1, 2, 5, 10, 20, 30 |
| | Epochs (w.r.t. ndays) | 50, 50, 70, 100, 200, 300 |

**4. Results and discussion**

4.1. Evaluation measures

4.1.1. Mean Absolute Percentage Error

Mean Absolute Percentage Error (MAPE) is often employed to assess the performance of the prediction methods. MAPE is also a measure of prediction accuracy for forecasting methods in machine learning area, it commonly presents accuracy as a percentage. Equation 2 shows its formula [30].

$$\text{MAPE} = \frac{1}{n}\sum_{t=1}^{n}\left|\frac{A_t - F_t}{A_t}\right| \times 100 \qquad (2)$$

where $A_t$ is the actual value and $F_t$ is the forecast value. In the formula, the absolute value of difference between those is divided by $A_t$. The absolute value is summed for every forecasted value and divided by the number of data. Finally, the percentage error is made by multiplying to 100.

4.1.2. Mean absolute error

Mean absolute error (MAE) is a measure of difference between two values. MAE is an average of the difference between the prediction and the actual values. MAE is a usual measure of prediction error for regression analysis in machine learning area. The formula is shown in Equation 3 [30].

$$\text{MAE} = \frac{1}{n}\sum_{t=1}^{n}|A_t - F_t| \qquad (3)$$

where $A_t$ is the true value and $F_t$ is the prediction value. In the formula, the absolute value of difference between those is divided by n (number of samples) and is summed for every forecasted value.

4.1.3. $R^2$

$R^2$ is known as R Squared or the determination coefficient, which reports the goodness of fit measure for prediction models. $R^2$ is a value between 0 (no-fit) and 1 (perfect fit) to present the variance proportion for a dependent parameter that is explained by an independent parameter in a regression analysis. It also indicates the relationship strength between an independent parameter and dependent one to examine how many of the observed variation can be clarified by the regression model's inputs. The formula is shown in Equation 4 [30].

$$R^2 = 1 - \frac{SS_{res}}{SS_{tot}} \qquad (4)$$

Where $SS_{res}$ and $SS_{tot}$ are Explained variation and Total variation respectively.

4.2 Results

Six tree-based models namely Decision Tree, Bagging, Random Forest, Adaboost, Gradient Boosting and XGBoost, and also three neural networks based algorithms (ANN, RNN and LSTM) are employed in prediction of the four stock market groups. For the purpose, prediction experiments for 1, 2, 5, 10, 15, 20 and 30 days in advance of time are conducted. Results for Diversified Financials, Petroleum, Non-metallic minerals and Basic metals are depicted in Tables 4-10, 11-17, 18-24 and 25-31 respectively. Moreover, the average performance of algorithms for each group is demonstrated in Tables 32-35.

It is prominent to note that comprehensive number of experiments are performed for each of the groups and prediction models with various model parameters. Following tables show the best parameters where minimum prediction error is obtained. Indeed, it is clear from the results that error values rise when prediction models are created for more and more number of days ahead. This may be evident for all algorithms.

Based on extensive experimental works and reported values the following results are obtained:
Among tree-based models
- Decision Tree always has the lowest rank for prediction
- For Diversified Financials and Petroleum groups, the best average performance belongs to Adaboost regressor

- For Non-metallic minerals and Basic metals, Gradient Boosting regressor has the best average performance
- XGboost is the best by considering accuracy, strength of fitting and running time all together

Through neural networks

- ANN generally occupies the bottom for forecasting
- LSTM models outperform RNN ones significantly

On the whole

LSTM is powerfully the best model for prediction all stock market groups with the lowest error and the best ability to fit, but the problem is the long run time

Table 4. Diversified Financials 1-Day ahead

| Prediction Models | Parameters | Error Measures | | |
|---|---|---|---|---|
| | | MAPE | MAE | $R^2$ |
| | ntrees | | | |
| Decision Tree | 1 | 1.29 | 23.05 | 0.9966 |
| Bagging | 400 | 0.92 | 15.80 | 0.9990 |
| Random Forest | 300 | 0.92 | 15.51 | 0.9991 |
| Adaboost | 250 | 0.91 | 15.09 | 0.9985 |
| Gradient Boosting | 300 | 1.02 | 19.19 | 0.9970 |
| XGBoost | 100 | 0.88 | 14.86 | 0.9994 |
| | epochs | | | |
| ANN | 1000 | 1.01 | 16.07 | 0.9992 |
| | ndays | | | |
| RNN | 1 | 1.77 | 20.20 | 0.9991 |
| LSTM | 5 | 0.54 | 6.02 | 0.9999 |

Table 5. Diversified Financials 2-Days ahead

| Prediction Models | Parameters | Error Measures | | |
|---|---|---|---|---|
| | | MAPE | MAE | $R^2$ |
| | ntrees | | | |
| Decision Tree | 1 | 1.52 | 25.93 | 0.9981 |
| Bagging | 150 | 1.11 | 18.31 | 0.9991 |
| Random Forest | 500 | 1.12 | 18.39 | 0.9991 |
| Adaboost | 400 | 1.11 | 19.56 | 0.9989 |
| Gradient Boosting | 300 | 1.14 | 19.51 | 0.9988 |
| XGBoost | 150 | 1.14 | 19.81 | 0.9989 |
| | epochs | | | |
| ANN | 1000 | 1.41 | 23.35 | 0.9983 |
| | ndays | | | |
| RNN | 10 | 1.95 | 16.98 | 0.9991 |
| LSTM | 2 | 0.43 | 4.46 | 1.0000 |

Table 6. Diversified Financials 5-Days ahead

| Prediction Models | Parameters | Error Measures | | |
|---|---|---|---|---|
| | | MAPE | MAE | $R^2$ |
| | ntrees | | | |
| Decision Tree | 1 | 1.66 | 28.94 | 0.9968 |
| Bagging | 150 | 1.45 | 24.00 | 0.9985 |
| Random Forest | 500 | 1.47 | 24.46 | 0.9984 |
| Adaboost | 400 | 1.39 | 23.91 | 0.9982 |
| Gradient Boosting | 350 | 1.35 | 24.05 | 0.9975 |
| XGBoost | 300 | 1.45 | 24.12 | 0.9986 |
| | epochs | | | |
| ANN | 1000 | 2.27 | 39.69 | 0.9951 |
| | ndays | | | |
| RNN | 10 | 1.91 | 14.75 | 0.9997 |
| LSTM | 30 | 0.75 | 5.21 | 1.0000 |

Table 7. Diversified Financials 10-Days ahead

| Prediction Models | Parameters | Error Measures | | |
|---|---|---|---|---|
| | | MAPE | MAE | $R^2$ |
| | ntrees | | | |
| Decision Tree | 1 | 2.09 | 34.00 | 0.9966 |
| Bagging | 250 | 1.88 | 31.47 | 0.9978 |
| Random Forest | 300 | 1.86 | 31.36 | 0.9978 |
| Adaboost | 200 | 1.58 | 25.63 | 0.9983 |
| Gradient Boosting | 500 | 1.74 | 28.00 | 0.9978 |
| XGBoost | 500 | 1.77 | 31.07 | 0.9976 |
| | epochs | | | |
| ANN | 1000 | 4.12 | 65.38 | 0.9875 |
| | ndays | | | |
| RNN | 5 | 1.66 | 15.21 | 0.9995 |
| LSTM | 10 | 0.57 | 6.84 | 0.9999 |

Table 8. Diversified Financials 15-Days ahead

| Prediction Models | Parameters | Error Measures | | |
|---|---|---|---|---|
| | | MAPE | MAE | $R^2$ |
| | ntrees | | | |
| Decision Tree | 1 | 2.28 | 41.29 | 0.9927 |
| Bagging | 100 | 2.24 | 37.61 | 0.9966 |
| Random Forest | 50 | 2.24 | 37.28 | 0.9966 |
| Adaboost | 300 | 1.83 | 28.83 | 0.9965 |
| Gradient Boosting | 200 | 1.97 | 35.95 | 0.9944 |
| XGBoost | 500 | 2.03 | 35.37 | 0.9964 |

| Prediction Models | Parameters | MAPE | MAE | R² |
|---|---|---|---|---|
| ANN | epochs 1000 | 5.05 | 85.46 | 0.9806 |
| RNN | ndays 10 | 1.95 | 19.47 | 0.9995 |
| LSTM | 20 | 0.77 | 10.03 | 0.9997 |

Table 9. Diversified Financials 20-Days ahead

| Prediction Models | Parameters | Error Measures | | |
|---|---|---|---|---|
| | | MAPE | MAE | R² |
| Decision Tree | ntrees 1 | 2.80 | 49.12 | 0.9913 |
| Bagging | 100 | 2.56 | 42.43 | 0.9962 |
| Random Forest | 450 | 2.57 | 42.66 | 0.9962 |
| Adaboost | 450 | 2.01 | 33.25 | 0.9969 |
| Gradient Boosting | 350 | 2.17 | 39.10 | 0.9947 |
| XGBoost | 500 | 2.30 | 39.30 | 0.9961 |
| ANN | epochs 1000 | 5.66 | 126.69 | 0.9658 |
| RNN | ndays 20 | 1.59 | 14.70 | 0.9997 |
| LSTM | 10 | 0.55 | 7.06 | 0.9999 |

Table 10. Diversified Financials 30-Days ahead

| Prediction Models | Parameters | Error Measures | | |
|---|---|---|---|---|
| | | MAPE | MAE | R² |
| Decision Tree | ntrees 1 | 2.83 | 48.39 | 0.9920 |
| Bagging | 350 | 3.21 | 54.37 | 0.9944 |
| Random Forest | 50 | 3.18 | 54.06 | 0.9944 |
| Adaboost | 350 | 2.33 | 37.63 | 0.9965 |
| Gradient Boosting | 500 | 2.54 | 43.59 | 0.9942 |
| XGBoost | 400 | 2.48 | 42.85 | 0.9961 |
| ANN | epochs 1000 | 7.48 | 126.69 | 0.9658 |
| RNN | ndays 20 | 2.11 | 19.09 | 0.9995 |
| LSTM | 10 | 0.61 | 7.25 | 0.9999 |

Table 11. Petroleum 1-Day ahead

| Prediction Models | Parameters | MAPE | MAE | $R^2$ |
|---|---|---|---|---|
| | ntrees | | | |
| Decision Tree | 1 | 1.72 | 4509.56 | 0.9987 |
| Bagging | 450 | 1.49 | 3777.47 | 0.9990 |
| Random Forest | 200 | 1.49 | 3778.05 | 0.9990 |
| Adaboost | 450 | 1.49 | 3758.63 | 0.9991 |
| Gradient Boosting | 450 | 1.43 | 3933.35 | 0.9989 |
| XGBoost | 150 | 1.39 | 3670.89 | 0.9991 |
| | epochs | | | |
| ANN | 1000 | 1.89 | 4424.08 | 0.9987 |
| | ndays | | | |
| RNN | 1 | 2.36 | 4566.10 | 0.9987 |
| LSTM | 2 | 1.20 | 1182.64 | 1.0000 |

Table 12. Petroleum 2-Days ahead

| Prediction Models | Parameters | MAPE | MAE | $R^2$ |
|---|---|---|---|---|
| | ntrees | | | |
| Decision Tree | 1 | 2.06 | 5741.50 | 0.9977 |
| Bagging | 200 | 1.75 | 4710.11 | 0.9986 |
| Random Forest | 100 | 1.74 | 4660.99 | 0.9986 |
| Adaboost | 100 | 1.71 | 4576.18 | 0.9987 |
| Gradient Boosting | 200 | 1.71 | 5064.24 | 0.9978 |
| XGBoost | 100 | 1.67 | 4516.52 | 0.9988 |
| | epochs | | | |
| ANN | 1000 | 2.47 | 6362.21 | 0.9974 |
| | ndays | | | |
| RNN | 1 | 3.46 | 6774.35 | 0.9973 |
| LSTM | 20 | 1.20 | 1989.73 | 0.9998 |

Table 13. Petroleum 5-Days ahead

| Prediction Models | Parameters | MAPE | MAE | $R^2$ |
|---|---|---|---|---|
| | ntrees | | | |
| Decision Tree | 1 | 2.48 | 6315.38 | 0.9971 |
| Bagging | 250 | 2.19 | 5011.67 | 0.9988 |
| Random Forest | 50 | 2.19 | 5109.34 | 0.9988 |
| Adaboost | 250 | 1.94 | 4312.98 | 0.9991 |
| Gradient Boosting | 250 | 1.87 | 4584.63 | 0.9985 |
| XGBoost | 500 | 2.03 | 4955.79 | 0.9987 |

| Prediction Models | Parameters | MAPE | MAE | R² |
|---|---|---|---|---|
| ANN | epochs 1000 | 3.68 | 9102.18 | 0.9948 |
| RNN | ndays 20 | 3.38 | 3081.28 | 0.9998 |
| LSTM | 30 | 1.50 | 1796.06 | 0.9999 |

Table 14. Petroleum 10- Days ahead

| Prediction Models | Parameters | Error Measures | | |
|---|---|---|---|---|
| | | MAPE | MAE | R² |
| Decision Tree | ntrees 1 | 2.71 | 7030.19 | 0.9955 |
| Bagging | 300 | 2.76 | 6554.66 | 0.9974 |
| Random Forest | 100 | 2.75 | 6563.85 | 0.9975 |
| Adaboost | 400 | 2.27 | 5082.99 | 0.9981 |
| Gradient Boosting | 350 | 2.31 | 6126.48 | 0.9968 |
| XGBoost | 500 | 2.53 | 6028.00 | 0.9978 |
| ANN | epochs 1000 | 5.05 | 13003.28 | 0.9892 |
| RNN | ndays 30 | 3.44 | 3086.82 | 0.9997 |
| LSTM | 5 | 1.19 | 1885.01 | 0.9999 |

Table 15. Petroleum 15- Days ahead

| Prediction Models | Parameters | Error Measures | | |
|---|---|---|---|---|
| | | MAPE | MAE | R² |
| Decision Tree | ntrees 1 | 2.91 | 8741.35 | 0.9919 |
| Bagging | 50 | 2.83 | 7041.03 | 0.9970 |
| Random Forest | 250 | 2.82 | 7026.39 | 0.9970 |
| Adaboost | 100 | 2.39 | 5668.14 | 0.9973 |
| Gradient Boosting | 400 | 2.37 | 7107.76 | 0.9935 |
| XGBoost | 500 | 2.55 | 6216.40 | 0.9979 |
| ANN | epochs 1000 | 6.75 | 16827.80 | 0.9820 |
| RNN | ndays 30 | 3.28 | 3656.84 | 0.9996 |
| LSTM | 10 | 1.03 | 1670.36 | 0.9999 |

Table 16. Petroleum 20- Days ahead

| Prediction Models | Parameters | Error Measures | | |
|---|---|---|---|---|
| | | MAPE | MAE | $R^2$ |
| | ntrees | | | |
| Decision Tree | 1 | 3.32 | 10006.94 | 0.9891 |
| Bagging | 150 | 3.33 | 9068.47 | 0.9950 |
| Random Forest | 450 | 3.34 | 9073.27 | 0.9953 |
| Adaboost | 50 | 2.64 | 6523.25 | 0.9952 |
| Gradient Boosting | 500 | 2.79 | 8157.44 | 0.9947 |
| XGBoost | 500 | 2.87 | 7862.15 | 0.9960 |
| | epochs | | | |
| ANN | 1000 | 7.85 | 20633.02 | 0.9754 |
| | ndays | | | |
| RNN | 20 | 3.92 | 3439.98 | 0.9997 |
| LSTM | 10 | 0.98 | 1806.04 | 0.9998 |

Table 17. Petroleum 30- Days ahead

| Prediction Models | Parameters | Error Measures | | |
|---|---|---|---|---|
| | | MAPE | MAE | $R^2$ |
| | ntrees | | | |
| Decision Tree | 1 | 3.72 | 10949.88 | 0.9828 |
| Bagging | 100 | 4.02 | 10319.46 | 0.9936 |
| Random Forest | 100 | 4.03 | 10332.38 | 0.9937 |
| Adaboost | 150 | 3.08 | 7031.89 | 0.9952 |
| Gradient Boosting | 450 | 3.35 | 9840.69 | 0.9910 |
| XGBoost | 400 | 3.26 | 8380.78 | 0.9953 |
| | epochs | | | |
| ANN | 1000 | 10.93 | 27967.86 | 0.9577 |
| | ndays | | | |
| RNN | 30 | 3.97 | 4075.02 | 0.9994 |
| LSTM | 2 | 1.19 | 1246.66 | 0.9999 |

Table 18. Non-metallic minerals 1-Day ahead

| Prediction Models | Parameters | Error Measures | | |
|---|---|---|---|---|
| | | MAPE | MAE | $R^2$ |
| | ntrees | | | |
| Decision Tree | 1 | 1.34 | 32.24 | 0.9991 |
| Bagging | 450 | 1.07 | 24.66 | 0.9994 |
| Random Forest | 150 | 1.07 | 24.59 | 0.9994 |
| Adaboost | 300 | 1.13 | 25.85 | 0.9993 |
| Gradient Boosting | 200 | 1.08 | 28.69 | 0.9989 |

| Prediction Models | Parameters | MAPE | MAE | R² |
|---|---|---|---|---|
| XGBoost | 450 | 1.09 | 24.44 | 0.9995 |
| ANN | epochs 1000 | 1.60 | 26.93 | 0.9990 |
| RNN | ndays 5 | 4.59 | 34.62 | 0.9996 |
| LSTM | 10 | 1.52 | 13.53 | 0.9999 |

Table 19. Non-metallic 2-Days ahead

| Prediction Models | Parameters | Error Measures | | |
|---|---|---|---|---|
| | | MAPE | MAE | R² |
| Decision Tree | ntrees 1 | 1.73 | 42.64 | 0.9983 |
| Bagging | 200 | 1.37 | 34.37 | 0.9990 |
| Random Forest | 400 | 1.37 | 34.27 | 0.9991 |
| Adaboost | 50 | 1.38 | 34.73 | 0.9987 |
| Gradient Boosting | 450 | 1.30 | 34.30 | 0.9988 |
| XGBoost | 150 | 1.35 | 32.01 | 0.9991 |
| ANN | epochs 1000 | 1.98 | 41.46 | 0.9979 |
| RNN | ndays 1 | 4.19 | 50.23 | 0.9981 |
| LSTM | 20 | 0.96 | 14.19 | 0.9998 |

Table 20. Non-metallic minerals 5-Days ahead

| Prediction Models | Parameters | Error Measures | | |
|---|---|---|---|---|
| | | MAPE | MAE | R² |
| Decision Tree | ntrees 1 | 1.88 | 45.77 | 0.9975 |
| Bagging | 50 | 1.66 | 35.89 | 0.9989 |
| Random Forest | 400 | 1.65 | 35.44 | 0.9989 |
| Adaboost | 400 | 1.58 | 34.17 | 0.9983 |
| Gradient Boosting | 150 | 1.48 | 35.43 | 0.9983 |
| XGBoost | 500 | 1.62 | 35.07 | 0.9991 |
| ANN | epochs 1000 | 3.58 | 66.00 | 0.9929 |
| RNN | ndays 1 | 4.40 | 73.47 | 0.9935 |
| LSTM | 5 | 1.75 | 17.36 | 0.9998 |

Table 21. Non-metallic minerals 10-Days ahead

| Prediction Models | Parameters | Error Measures | | |
|---|---|---|---|---|
| | | MAPE | MAE | $R^2$ |
| | ntrees | | | |
| Decision Tree | 1 | 2.38 | 58.35 | 0.9954 |
| Bagging | 150 | 2.11 | 47.71 | 0.9981 |
| Random Forest | 450 | 2.11 | 47.84 | 0.9981 |
| Adaboost | 450 | 1.90 | 39.07 | 0.9989 |
| Gradient Boosting | 450 | 1.88 | 47.16 | 0.9969 |
| XGBoost | 400 | 1.98 | 43.20 | 0.9988 |
| | epochs | | | |
| ANN | 1000 | 4.18 | 100.80 | 0.9841 |
| | ndays | | | |
| RNN | 20 | 6.77 | 39.40 | 0.9995 |
| LSTM | 30 | 2.41 | 21.34 | 0.9997 |

Table 22. Non-metallic minerals 15-Days ahead

| Prediction Models | Parameters | Error Measures | | |
|---|---|---|---|---|
| | | MAPE | MAE | $R^2$ |
| | ntrees | | | |
| Decision Tree | 1 | 2.47 | 56.71 | 0.9962 |
| Bagging | 150 | 2.42 | 55.74 | 0.9978 |
| Random Forest | 150 | 2.43 | 56.60 | 0.9978 |
| Adaboost | 250 | 2.04 | 45.75 | 0.9984 |
| Gradient Boosting | 250 | 2.03 | 47.49 | 0.9977 |
| XGBoost | 500 | 2.17 | 50.03 | 0.9984 |
| | epochs | | | |
| ANN | 1000 | 6.10 | 129.67 | 0.9809 |
| | ndays | | | |
| RNN | 5 | 5.59 | 30.61 | 0.9997 |
| LSTM | 20 | 1.60 | 23.18 | 0.9996 |

Table 23. Non-metallic minerals 20-Days ahead

| Prediction Models | Parameters | Error Measures | | |
|---|---|---|---|---|
| | | MAPE | MAE | $R^2$ |
| | ntrees | | | |
| Decision Tree | 1 | 2.47 | 59.36 | 0.9971 |
| Bagging | 350 | 2.67 | 58.48 | 0.9978 |
| Random Forest | 250 | 2.66 | 58.52 | 0.9978 |
| Adaboost | 200 | 2.14 | 48.71 | 0.9984 |
| Gradient Boosting | 400 | 2.08 | 50.78 | 0.9978 |
| XGBoost | 500 | 2.19 | 49.64 | 0.9979 |
| ANN | epochs 1000 | 6.99 | 158.87 | 0.9691 |
| RNN | ndays 30 | 6.74 | 49.13 | 0.9991 |
| LSTM | 20 | 1.19 | 14.02 | 0.9999 |

Table 24. Non-metallic minerals 30-Days ahead

| Prediction Models | Parameters | Error Measures | | |
|---|---|---|---|---|
| | | MAPE | MAE | $R^2$ |
| Decision Tree | ntrees 1 | 3.02 | 74.19 | 0.9918 |
| Bagging | 100 | 3.53 | 78.32 | 0.9946 |
| Random Forest | 350 | 3.52 | 77.98 | 0.9944 |
| Adaboost | 400 | 2.70 | 60.86 | 0.9936 |
| Gradient Boosting | 450 | 2.58 | 58.99 | 0.9948 |
| XGBoost | 500 | 2.65 | 60.66 | 0.9954 |
| ANN | epochs 1000 | 8.28 | 178.22 | 0.9691 |
| RNN | ndays 10 | 4.32 | 31.77 | 0.9994 |
| LSTM | 20 | 1.24 | 14.94 | 0.9998 |

Table 25. Metals 1-Day ahead

| Prediction Models | Parameters | Error Measures | | |
|---|---|---|---|---|
| | | MAPE | MAE | $R^2$ |
| Decision Tree | ntrees 1 | 0.90 | 734.48 | 0.9991 |
| Bagging | 400 | 0.71 | 581.16 | 0.9994 |
| Random Forest | 150 | 0.72 | 590.17 | 0.9994 |
| Adaboost | 400 | 0.75 | 608.28 | 0.9993 |
| Gradient Boosting | 200 | 0.74 | 643.71 | 0.9991 |
| XGBoost | 200 | 0.72 | 574.99 | 0.9995 |
| ANN | epochs 1000 | 0.91 | 608.74 | 0.9995 |
| RNN | nDays 1 | 1.27 | 689.38 | 0.9994 |
| LSTM | 2 | 0.68 | 352.85 | 0.9999 |

Table 26. Metals 2-Days ahead

| Prediction Models | Parameters | Error Measures | | |
|---|---|---|---|---|
| | | MAPE | MAE | $R^2$ |
| | ntrees | | | |

| Prediction Models | Parameters | MAPE | MAE | R² |
|---|---|---|---|---|
| Decision Tree | 1 | 1.00 | 835.22 | 0.9990 |
| Bagging | 250 | 0.81 | 660.18 | 0.9994 |
| Random Forest | 300 | 0.81 | 666.72 | 0.9994 |
| Adaboost | 350 | 0.84 | 698.59 | 0.9993 |
| Gradient Boosting | 400 | 0.81 | 691.07 | 0.9992 |
| XGBoost | 450 | 0.80 | 669.77 | 0.9995 |
| ANN | epochs 1000 | 1.22 | 930.33 | 0.9989 |
| RNN | nDays 1 | 1.36 | 972.59 | 0.9989 |
| LSTM | 5 | 0.73 | 257.32 | 1.0000 |

Table 27. Metals 5-Days ahead

| Prediction Models | Parameters | Error Measures | | |
|---|---|---|---|---|
| | | MAPE | MAE | R² |
| Decision Tree | ntrees 1 | 1.18 | 916.51 | 0.9985 |
| Bagging | 450 | 1.07 | 830.27 | 0.9991 |
| Random Forest | 300 | 1.06 | 819.80 | 0.9991 |
| Adaboost | 400 | 0.97 | 714.24 | 0.9993 |
| Gradient Boosting | 450 | 0.92 | 710.35 | 0.9992 |
| XGBoost | 500 | 1.04 | 796.40 | 0.9992 |
| ANN | epochs 1000 | 1.98 | 1476.84 | 0.9975 |
| RNN | nDays 10 | 1.89 | 634.63 | 0.9998 |
| LSTM | 20 | 0.37 | 188.99 | 1.0000 |

Table 28. Metals 10-Days ahead

| Prediction Models | Parameters | Error Measures | | |
|---|---|---|---|---|
| | | MAPE | MAE | R² |
| Decision Tree | ntrees 1 | 1.32 | 1004.13 | 0.9986 |
| Bagging | 200 | 1.33 | 988.90 | 0.9991 |
| Random Forest | 150 | 1.32 | 987.82 | 0.9991 |
| Adaboost | 300 | 1.15 | 836.12 | 0.9993 |
| Gradient Boosting | 350 | 1.10 | 902.71 | 0.9989 |
| XGBoost | 500 | 1.21 | 952.97 | 0.9990 |
| ANN | epochs 1000 | 3.12 | 2335.23 | 0.9940 |
| RNN | nDays 30 | 0.95 | 448.31 | 0.9999 |
| LSTM | 2 | 0.31 | 189.37 | 1.0000 |

Table 29. Metals 15-Days ahead

| Prediction Models | Parameters | Error Measures | | |
|---|---|---|---|---|
| | | MAPE | MAE | R² |
| | ntrees | | | |
| Decision Tree | 1 | 1.64 | 1388.14 | 0.9957 |
| Bagging | 350 | 1.67 | 1293.94 | 0.9982 |
| Random Forest | 300 | 1.67 | 1290.31 | 0.9982 |
| Adaboost | 500 | 1.42 | 1031.74 | 0.9989 |
| Gradient Boosting | 250 | 1.40 | 1187.28 | 0.9965 |
| XGBoost | 250 | 1.51 | 1232.24 | 0.9979 |
| | epochs | | | |
| ANN | 1000 | 3.86 | 3053.19 | 0.9897 |
| | nDays | | | |
| RNN | 30 | 1.88 | 703.94 | 0.9997 |
| LSTM | 30 | 0.52 | 338.03 | 0.9999 |

Table 30. Metals 20-Days ahead

| Prediction Models | Parameters | Error Measures | | |
|---|---|---|---|---|
| | | MAPE | MAE | R² |
| | ntrees | | | |
| Decision Tree | 1 | 1.89 | 1610.56 | 0.9957 |
| Bagging | 400 | 1.95 | 1532.22 | 0.9976 |
| Random Forest | 450 | 1.94 | 1517.01 | 0.9975 |
| Adaboost | 450 | 1.56 | 1113.56 | 0.9988 |
| Gradient Boosting | 150 | 1.55 | 1266.97 | 0.9971 |
| XGBoost | 500 | 1.61 | 1295.44 | 0.9983 |
| | epochs | | | |
| ANN | 1000 | 4.74 | 3862.35 | 0.9857 |
| | nDays | | | |
| RNN | 20 | 1.65 | 628.05 | 0.9998 |
| LSTM | 30 | 0.56 | 354.33 | 0.9999 |

Table 31. Metals 30-Days ahead

| Prediction Models | Parameters | Error Measures | | |
|---|---|---|---|---|
| | | MAP | MAE | R² |
| | ntrees | | | |
| Decision Tree | 1 | 1.95 | 1627.18 | 0.9936 |
| Bagging | 450 | 1.99 | 1439.79 | 0.9976 |
| Random Forest | 450 | 1.99 | 1431.28 | 0.9977 |
| Adaboost | 400 | 1.56 | 1036.61 | 0.9988 |
| Gradient Boosting | 400 | 1.59 | 1321.55 | 0.9970 |
| XGBoost | 500 epochs | 1.61 | 1222.12 | 0.9983 |
| ANN | 1000 nDays | 6.39 | 4825.30 | 0.9794 |
| RNN | 30 | 1.38 | 567.24 | 0.9998 |
| LSTM | 2 | 0.59 | 229.76 | 1.0000 |

Table 32. Average performance for Diversified Financials

| Prediction Models | Error Measures | | |
|---|---|---|---|
| | MAP | MAE | $R^2$ |
| | | | |
| Decision Tree | 2.07 | 35.82 | 0.9949 |
| Bagging | 1.91 | 32.00 | 0.9974 |
| Random Forest | 1.91 | 31.96 | 0.9974 |
| Adaboost | 1.59 | 26.27 | 0.9977 |
| Gradient Boosting | 1.70 | 29.91 | 0.9963 |
| XGBoost | 1.72 | 29.63 | 0.9976 |
| | | | |
| ANN | 3.86 | 69.05 | 0.9846 |
| | | | |
| RNN | 1.85 | 17.20 | 0.9994 |
| LSTM | 0.60 | 6.70 | 0.9999 |

Table 33. Average performance for Petroleum

| Prediction Models | Error Measures | | |
|---|---|---|---|
| | MAP | MAE | $R^2$ |
| | | | |
| Decision Tree | 2.70 | 7613.54 | 0.9933 |
| Bagging | 2.62 | 6640.41 | 0.9971 |
| Random Forest | 2.62 | 6649.18 | 0.9971 |
| Adaboost | 2.22 | 5279.15 | 0.9975 |
| Gradient Boosting | 2.26 | 6402.08 | 0.9959 |
| XGBoost | 2.33 | 5947.22 | 0.9977 |
| | | | |
| ANN | 5.52 | 14045.78 | 0.9850 |
| | | | |
| RNN | 3.40 | 4097.20 | 0.9992 |
| LSTM | 1.18 | 1653.79 | 0.9999 |

Table 34. Average performance for Non-metallic minerals

| Prediction Models | Error Measures | | |
|---|---|---|---|
| | MAP | MAE | $R^2$ |
| Decision Tree | 2.18 | 52.75 | 0.9965 |
| Bagging | 2.12 | 47.88 | 0.9979 |
| Random Forest | 2.12 | 47.89 | 0.9979 |
| Adaboost | 1.84 | 41.31 | 0.9979 |
| Gradient Boosting | 1.78 | 43.26 | 0.9976 |
| XGBoost | 1.86 | 42.15 | 0.9983 |
| ANN | 4.67 | 100.28 | 0.9847 |
| RNN | 5.23 | 44.18 | 0.9984 |
| LSTM | 1.52 | 16.94 | 0.9998 |

Table 35. Average performance for Metals

| Prediction Models | Error Measures | | |
|---|---|---|---|
| | MAP | MAE | $R^2$ |
| Decision Tree | 1.41 | 1159.46 | 0.9972 |
| Bagging | 1.36 | 1046.64 | 0.9986 |
| Random Forest | 1.36 | 1043.30 | 0.9986 |
| Adaboost | 1.18 | 862.73 | 0.9991 |
| Gradient Boosting | 1.16 | 960.52 | 0.9981 |
| XGBoost | 1.21 | 963.42 | 0.9988 |
| ANN | 3.17 | 2441.71 | 0.9921 |
| RNN | 1.48 | 663.45 | 0.9996 |
| LSTM | 0.54 | 272.95 | 1.0000 |

## 5. Conclusion

For all investors it is always necessary to predict stock market changes for detecting accurate profits and reducing potential mark risks. This study effort was employing Tree-based models (Decision Tree, Bagging, Random Forest, Adaboost, Gradient Boosting and XGBoost) and neural networks (ANN, RNN and LSTM) in order to correctly forecast the values of four stock market groups (Diversified Financials, Petroleum, Non-metallic minerals and Basic metals) as a regression problem. The predictions were made for 1, 2, 5, 10, 15, 20 and 30 days ahead. As far as our belief and knowledge, this study is the successful and recent research work that involves ensemble learning methods and deep learning algorithms for predicting stock groups as a popular application. To be more detailed, exponentially smoothed technical indicators and features were used as inputs for prediction. In this prediction problem, the methods were able to significantly advance their performance, and LSTM was the top performer in comparison with other techniques. Overall, as a logical conclusion, both tree-based and deep learning algorithms showed remarkable potential in regression problems in the area of machine learning.

**Author contributions:** Data curation, Mojtaba Nabipour, Pooyan Nayyeri, Hamed; Formal analysis, Mojtaba Nabipour, Pooyan Nayyeri, Hamed Jabani; Funding acquisition, Amir Mosavi; Investigation, Pooyan Nayyeri and Hamed Jabani; Project administration, Amir Mosavi; Resources, Mojtaba Nabipour; Software, Mojtaba Nabipour, Pooyan Nayyeri; Supervision, Amir Mosavi; Visualization, Hamed Jabani; Writing – original draft, Mojtaba Nabipour; Writing – review & editing, Amir Mosavi; Conceptualization, Amir Mosavi.